\documentclass[twocolumn,reprint,10pt]{revtex4}
\usepackage{epsfig}
\usepackage{bm}
\usepackage{color}
\usepackage {amsmath}
\DeclareGraphicsExtensions{.pdf,.png,.jpg}

\newcommand{\sss}{\scriptscriptstyle}
\newcommand{\s}{\scriptstyle}
\begin{document}

\title{Anisotropic relaxation in NADH excited states studied by polarization-modulation pump-probe transient spectroscopy}
\author{Ioanna A. Gorbunova, Maxim E. Sasin, Yaroslav M. Beltukov, Alexander A. Semenov, and Oleg S. Vasyutinskii}
\affiliation{Ioffe Institute, Polytekhnicheskaya 26, St.Petersburg, Russia}

\date{May 5, 2020 }

\begin{abstract}
We present the results of experimental and theoretical studies of fast anisotropic relaxation and rotational diffusion in the first electron excited state of biological coenzyme NADH in water-ethanol solutions. The experiments have been carried out by means of a novel polarization-modulation transient method and fluorescence polarization spectroscopy. For interpretation of the experimental results a  model of the anisotropic relaxation in terms of scalar and vector properties of transition dipole moments and based on the Born-Oppenheimer approximation has been developed. The results obtained suggest that the dynamics of anisotropic rovibronic relaxation in NADH under excitation with 100~fs pump laser pulses can be characterised by a single vibration relaxation time $\tau_v$ laying in the range 2--15~ps and a single rotation diffusion time $\tau_r$ laying in the range 100--450~ps a subject of ethanol concentration. The dependence of the times $\tau_v$ and $\tau_r$ on the solution polarity (static permittivity) and viscosity has been determined and analyzed. Limiting values of an important parameter $\langle P_2(\cos\theta(t))\rangle$  describing the rotation of the transition dipole moment in the course of vibrational relaxation has been determined from experiment as function of the ethanol concentration and analyzed.
\end{abstract}

\maketitle

\section{Introduction}
Transient absorption spectroscopy is widely used nowadays as a powerful tool for investigation of ultrafast photoinduced processes of electron and proton transfer, isomerization, and excited state dynamics. These studies have received considerable interest within broad scientific community because the dynamics of elementary physical and chemical processes in the gas and condensed phases can be followed in real time by femtosecond pump-probe spectroscopy \cite{Mukamel95,Berera09,Henriksen14,Zhu15,Fisher16}. Ultrafast transient absorption spectroscopy of biologically relevant molecules allows to reach a previously inaccessible level of understanding of dynamics of intra- and intermolecular interactions including relaxation in the ground and excited states, nonradiative recombination, electron and energy transfer, etc \cite{Fleming96,Cohen03,Ewing14,Heiner13,Heiner17}.

 %The pump-probe teсhnic, where  pump pulse induces  photoprocess, and  probe pulse delayed relative to probe, controls change in the optical %properties of the sample, proved to be especially effective.

%Pump-probe spectroscopy has been used for many years to study fast photoinduced processes in gas and condensed matter.

Excited state dynamics in biologically relevant molecules is closely related to fast anisotropic relaxation and rotation diffusion processes, that can be successfully studied by multiphoton polarized fluorescence spectroscopy \cite{Lakowicz97a,Couprie94,Vishwasrao06,Denicke10,Blacker13,Herbrich15,Sasin18}, or by polarization-resolved pump-probe ultrafast spectroscopy \cite{Levenson79,Hochstrasser96,Beck98,Jonas01,Wiersma01,Fayer05,Jonas08,Fenn09,Jonas11,Tros15,Corrales16,Rumble19,Hunger19}. In both cases a linearly polarized pump pulse resonantly promotes ground state molecules to an excited electronic/vibrational state thus producing an anisotropic distribution of excited molecular axes/bonds that becomes isotropic as a function of time due to random orientational motion of the molecules and relaxation processes. Within the pumpe-probe procedure a real time anisotropy decay is measured by absorption change of a delayed probe pulse that is polarized either parallel, or perpendicularly to the pump pulse:
\begin{equation}
\label{eq:Idiff1}
R(t)=\frac{I_{\|}(t)-I_{\bot}(t)}{I_{\|}(t)+2I_{\bot}(t)},
\end{equation}
where $R(t)$ is the anisotropy and $t$ is a delay time between the pump and probe pulses. The nominator in eq.(\ref{eq:Idiff1}) is the linear absorption dichroism of the probe beam.

Anisotropic relaxation and rotation diffusion decay times that can be determined in these experiments are of great importance for understanding of complicated biochemical processes in living cells and tissues. For instance, the rotation diffusion times $\tau_{r}$ under certain conditions are  proportional to  solution viscosity~\cite{ZEWAIL79} and therefore can be used for determination of local intracellular and extracellular viscosity~\cite{Wiersma01}. Moreover, the decay times and other fluorescence/probe signal parameters depend in general on the location of a fluorophore inside or outside cells and on intracellular processes occurring and therefore can be used as noninvasive markers of cellular energy metabolism \cite{Schaefer19}.

An important problem in recording of the excited state absorption dichroism $\Delta I_{ab}(t)=I_{\|}(t)-I_{\bot}(t)$ in eq.(\ref{eq:Idiff1}) is a relatively small signal-to-noise ratio because transient probe beam intensity $I_{pr}$ transmitting through the experimental sample and pointing a photodetector is usually much larger than each of the absorption intensities $I_{\|}(t)$ and $I_{\bot}(t)$ and moreover than their difference $\Delta I_{ab}(t)$. Due to relatively high amplitude fluctuations of a femtosecond laser output the noise in a recording system caused by the probe beam detection can become comparable with the absorption intensities and even larger.

As absorption signal is proportional to the absorbing molecules concentration and to the laser pulse energies, the major number of the transient spectroscopy experiments have been carried out till now either used relatively high energy laser pulses of several tens, or hundreds of $nJ$ and more (see, e.g. \cite{Hochstrasser96,Ewing14,Corrales16,Hunger19}, or probed condense state or ground state molecules~\cite{Fleming96,Beck98,Jonas01,Wiersma01,Jonas08,Fenn09,Tros15}.

As pointed out by Jonas et al~\cite{Fleming96} a low pump and probe pulse energies (typically less than a $nJ$) are required in experiments with biological samples to prevent sample bleaching and damaging by laser light. The problem of decreasing of signal-to-noise ratio was solved  by several groups for various pump-probe experimental schemes.

Jonas et al.~\cite{Fleming96} used a balanced detection scheme included probe and reference beam channels combined with the pump beam amplitude modulation by a chopper and detection of the modulated transient signal with a lock-in detector. Fenn et al \cite{Fenn09} successively measured $I_{\|}(t)$ and $I_{\bot}(t)$ using a polarizer placed in the probe beam, and averaged many successive parallel and perpendicular measurements. Tros et al \cite{Tros15} suggested an improvement of the approach \cite{Fenn09} by measuring  $I_{\|}(t)$ and $I_{\bot}(t)$  alternately on a shot-to-shot basis. In this way, the IR fluctuations at much smaller frequencies than the pulse repetition rate (about 1~kHz \cite{Tros15}) effectively canceled out, resulting in a better signal-to-noise ratio. However, in that case a higher frequency noise still contributes to the measured anisotropy in eq.(\ref{eq:Idiff1}).

\begin{figure}[h!]\center
\includegraphics[width=0.3\textwidth]{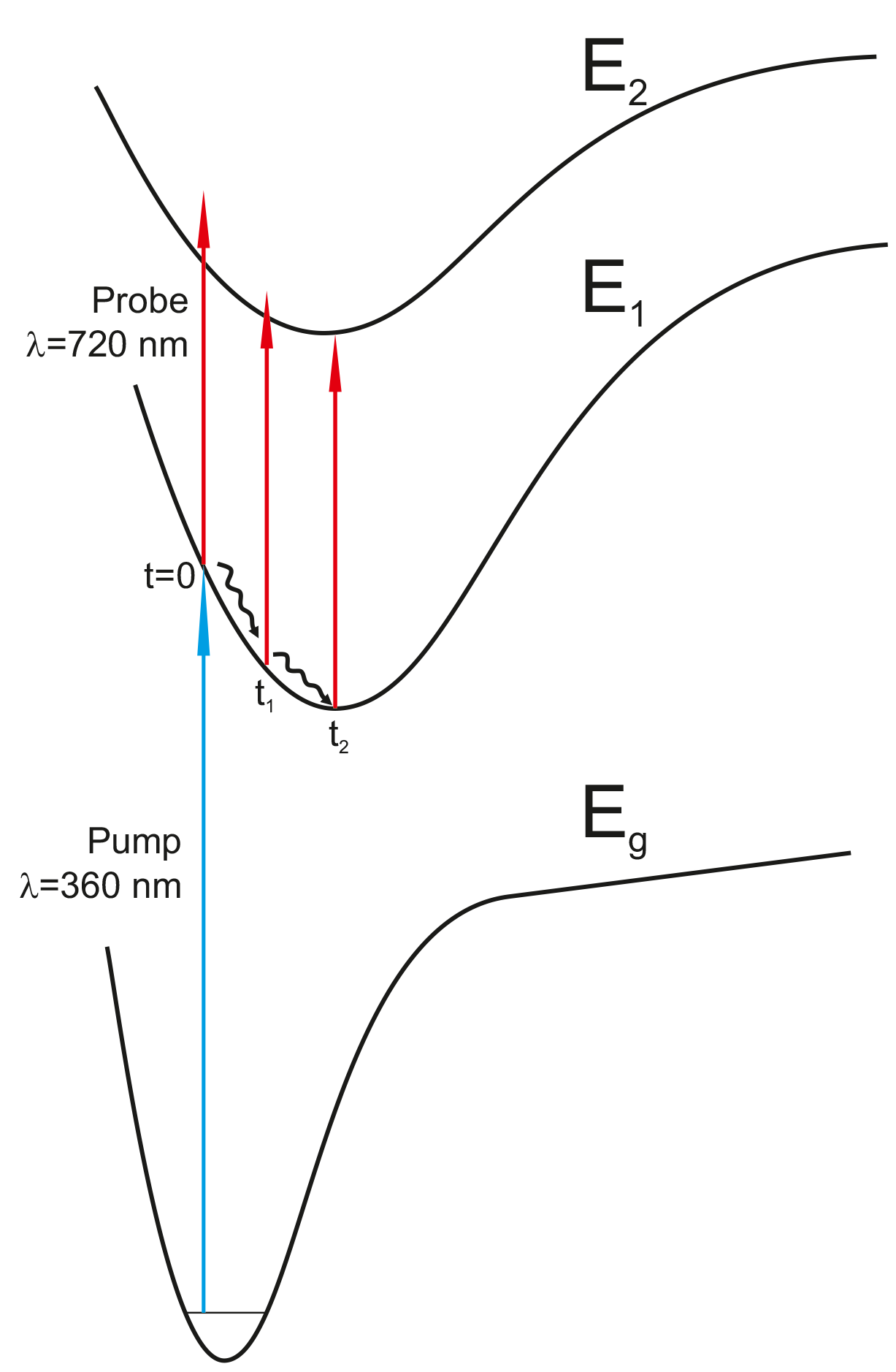}
\caption{Schematic of the resonant two-step excitation}
\label{fig:resonant}
\end{figure}

This paper presents the results of our study of anisotropic relaxation in the first excited state of biological coenzyme NADH (reduced nicotinamide adenine dinucleotide) in ethanol-water solutions by means of fluorescence polarization spectroscopy and a novel polarization-modulation transient method. The method utilizes the geometry of the optical Kerr effect (OKE) experiments~\cite{Levenson79,Wiersma01} where the pump and probe beams pass collinearly through the sample with their linear polarizations directed at 45$^\circ$ to each other. Unlike most of the OKE studies reported till now in our experiment both pump and probe beams optical frequencies were in resonance with molecular transitions and therefore the interpretation of the results obtained could be done in a way similar to that used for conventional pump-probe polarization experiments~\cite{Fleming96}. Linearly polarized pump beam resonantly promoted ground state NADH molecules to the first electronic excited state and produced the excited state axes/bonds alignment that was detected by recording of the absorption linear dichroism $\Delta I_{ab}(t)$ of the probe beam on the optical transition from the first to the second excited state as shown in Fig.\ref{fig:resonant}.
The method significantly enhances the accuracy of  transient polarization-sensitive measurements and can be used for investigation of excited state dynamics in biological molecules at a less than nJ level of pump pulse energy.

Coenzyme NADH is one of the most important biological molecules, since it plays an active role in redox reactions in living cells. Possessing intense fluorescence NADH is widely used as a natural fluorescence marker for studying biochemical processes in living cells.  Up to now, a number of time-resolved fluorescence studies have been carried out in NADH  in solutions and cells (see e.g. \cite{Freed67,Hull01,Vishwasrao06,Blacker13,Sasin19,Vasyutinskii17,Blacker19,Schaefer19}).  NADH has complicated excited state photodynamics possessing two chromophore groups: nicotinamide and adenine that can interact with each other under certain conditions \cite{Schaefer19}. This dynamics was investigated recently by Heiner et al \cite{Heiner13,Heiner17} using the transient absorption spectroscopy.

However, despite of numerous studies, energy transfer processes in the excited states of NADH are currently far from clear understanding.  To the best of our knowledge no transient polarization-sensitive studies have been carried out till now.

In this paper the fluorescence polarization spectroscopy and the polarization-modulation transient method were used for studying of fast anisotropic relaxation and rotational diffusion in the first electronic excited state of biological coenzyme NADH (nicotinamide-adenine-dinucleotide) in aqueous-ethanol solutions in a temporal range from sub-picoseconds to several nanoseconds. The results obtained suggest that anisotropic ro-vibronic relaxation under excitation with 100~fs pump laser pulses can be characterised by a single vibration relaxation time $\tau_v$ laying in the range 2--15~ps and a single rotation diffusion time $\tau_{r}$ laying in the range 100--450~ps a subject of ethanol concentration.  A model of the anisotropic relaxation in terms of scalar and vector properties of transition dipole moments and based on the Born-Oppenheimer approximation has been developed.

The paper is organised as follows. Section II describes the quantum mechanics-based expressions for experimental signals that justify the physical interpretation of the polarization-modulation transient method developed. Section III describes the details of the experimental procedure used. The experimental results obtained are collected in Sec. IV. Discussion of the results obtained and their interpretation are given in Section V. The conclusions are given in Sec. VI. The calculation details are collected in Appendix.

\section{Theory}
\label{sec:theory}
An expression describing the transient polarization signals observed in our experiments is based on the excitation scheme shown in Fig.\ref{fig:resonant} suggesting that linearly polarized pump beam resonantly promotes ground state molecules to the first electronic excited state and produces the excited state axes/bonds alignment that is detected by recording of the probe beam absorption linear dichroism $\Delta I_{ab}(t)$ on resonant transitions from the first to the second excited state. In the case of NADH, the first excited state refers to the excitation of the nicotinamide chromophore group via the absorption band centered at about 340~nm and the second excited state refers to the excitation of the adenine chromophore group via the absorption band centered at about 260~nm.
\begin{figure}[h!]\center
\includegraphics[width=0.4\textwidth]{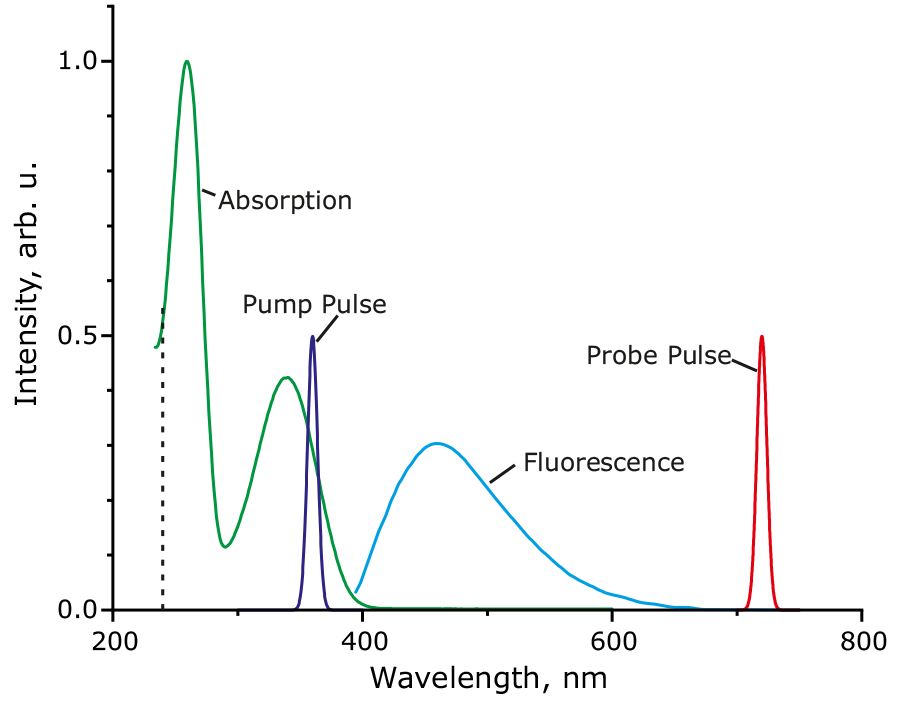}
\caption{Absorption and emission spectral bands of NADH. Pump and probe pulse spectral profiles are centered at 360~nm and 720~nm, respectively.}
\label{fig:spectra}
\end{figure}
Absorption and emission spectra of NADH are given in Fig.2. The fluorescence band in this Fig.2 centered at about 460~nm occurs due to spontaneous emission from the excited nicotinamide group to the ground electronic state~\cite{Schaefer19}. Sharp spectral profiles at 360 and 720~nm  in Fig.2 represent the pump and probe pulses, respectively and the vertical dashed line indicates the maximum available energy achieved by two-photon excitation from the ground state at the delay time $t=0$.  At the delay times $t>0$ the wave packet in the first excited state moves down toward the equilibrium nuclear configuration point due to vibration relaxation processes as schematically shown in Fig.1 and the available excitation energy is shifted down below the dashed line in Fig.2.

According to the analysis of Jonas et al~\cite{Fleming96} the signal in pump-probe spectroscopy can be presented as a sum of signals
from three sources: signals from ground state depletion, signals from excited state stimulated emission, and signals from excited state absorption. As can be shown in the conditions of our experiments only the signals from excited absorption contribute to the experimental signals. Indeed, one can conclude from Fig.2 that the probe beam at 720~nm does not have enough energy to excite the ground state molecules alone and therefore, could not probe the ground state depletion at the times when the pump and probe pulses do not overlap. Also, as can be seen in Fig.2 the probe beam locates well outside of the fluorescence band and therefore contribution from excited state stimulated emission can also be not accounted for.

Therefore, the theoretical treatment presented below takes into account only probe beam absorption from the first to the second excited states in Fig.1. The theory is restricted to the relatively large delay times $t$ when overlap between the pump and probe pulses can be neglected.

The proposed theoretical model is essentially quantum mechanical and based on the spherical tensors approach~\cite{Blum96}.  A general case of excitation of an ensemble of asymmetric top molecules by two laser pulses has been considered.

We denote $J_g,J_1,J_2$ and $M_g,M_1,M_2$ the molecular total angular momenta of the ground, first, second excited states and their projections onto the laboratory axis Z, respectively.  Interaction between the molecular electric dipole moment and the incident pump and probe laser beams can be written in a   usual form as scalar products: $V_{pu} \sim \,\mathbf{d}^{(pu)}\mathbf{e}_{pu}$, $V_{pr} \sim \,\mathbf{d}^{(pr)}\mathbf{e}_{pr}$, where $\mathbf{d}^{(pu)}$ and $\mathbf{d}^{(pr)}$ are ground and excited state molecular dipole moments, and $\mathbf{e}_{pu}$ and $\mathbf{e}_{pr}$ are pump and probe light polarization vectors, respectively.

In the Born-Oppenheimer approximation the molecular wave functions can be presented as a product of the electronic $\psi^{el}_i$, vibration $\chi_i$, and rotation $\Psi^{rot}$ wave functions:
\begin{equation}
\label{eq:BO approximation1}
     |n_i J_i M_i\rangle = |n_i \rangle \Psi^{rot}_{J_i,M_i,k_i}(\alpha,\beta,\gamma),
\end{equation}
\begin{equation}
\label{eq:BO approximation2}
     |n_i \rangle = \psi^{el}_i\chi_i,
\end{equation}
where $i=g, 1,2$ and
\begin{eqnarray}
\label{E:A}
\Psi^{rot}_{J_i,M_i,k_i}(\alpha,\beta,\gamma)  = \sum_{\Omega_i}A^{J_i}_{k_i \Omega_i} D^{{J_i}^{\s *}}_{M_i,\Omega_i}(\alpha,\beta,\gamma)
\end{eqnarray}
are asymmetric top rotation wave functions~\cite{Zare88b} where $D^{J_i}_{M_i,\Omega_i}(\alpha,\beta,\gamma)$ is a Wigner $D$-function, the Euler angles $(\alpha,\beta,\gamma)$ specify the direction of a molecular general axis, $\Omega_i$ is a body frame projection of the angular momentum $J_i$, and $A^{J_i}_{k_i \Omega_i}$ is an expansion coefficient.

Combining the density matrix expressions for single photon transitions~\cite{Blum96} $g \rightarrow 1$ and $1 \rightarrow 2$  after some transformations of the angular momentum algebra  the expression for the probe beam absorption can be presented in the form (see Appendix~\ref{app:A}):
\begin{widetext}
\begin{eqnarray}
\label{eq:probe2}
    I_{ab}(t)&=& C_{pu}I_{pr}\sum_l\,a_l\,e^{-t/\tau_l}
   \sum_{ K} \Big(E_{K}(\textbf{e}_{pu})\cdot E_{K}(\textbf{e}_{pr})\Big)
                  \sum_{\tilde q, \tilde q'}D^K_{\tilde q' \tilde q}(t)
                   \sum_{q_1,^{\vphantom{\prime}}q_1'}\sum_{q_2^{\vphantom{\prime}},q_2'}C^{1 q_1' }_{1 q_1 \: K -\tilde q'}\, C^{1 q_2' }_{1 q_2 \: K \tilde q}
            \nonumber \\
    &\times&
     \langle n_2 |d^{(pr)}_{q_2'}| n'_{1rel} \rangle\langle n'_1 |d^{(pu)}_{q_1'}| n_g \rangle
     \langle n_2 |d^{(pr)}_{q_2^{\vphantom{\prime}}}| n_{1rel} \rangle^*\langle n_1 |d^{(pu)}_{q_1^{\vphantom{\prime}}}| n_g \rangle^*\, N_g,
   \end{eqnarray}
   \end{widetext}
where $C_{pu}$ is a constant proportional to the pump beam intensity, $\tau_l$ is a first excited state lifetime, $a_l$ is a weighting coefficient, $d_q$ is a spherical component of the molecular dipole moment onto the body axis with $q=0,\pm1$, and $C^{1 q_1' }_{1 q_1 \: K -\tilde q'}$ is a Clebsch-Gordan coefficient~\cite{Zare88b}.

The term $N_g$ in eq.(\ref{eq:probe2}) is the ground state molecular population defined as:
 \begin{equation}
\label{E:Ng}
   N_g= \sum_{J_g,k_g,\Omega_g}\,(2J_g+1)\,N(J_g,k_g){A^{J_g}_{k_g \Omega_g}}^{\s *}{A^{J_g}_{k_g \Omega_g}}\,
\end{equation}

The term in parenthesis in the first line in eq.(\ref{eq:probe2}) is a scalar product, where $E_{KQ}(\textbf{e})$ are light polarization density matrices \cite{Zare88b} of the pump and probe light:
\begin{equation}
\label{E:Edef}
    E_{KQ}(\textbf{e})=\left[\,\mathbf{e}\otimes\mathbf{e}^*\right]_{KQ}=
    \sum_{\mu \mu'}(-1)^{\mu'} C^{KQ}_{1\mu\,1-\mu'}\, {e}_{\mu}\,{e}_{\mu'}^*,
\end{equation}

The function $D^K_{\tilde q' \tilde q}(t)$ in eq.(\ref{eq:probe2}) describes rotation diffusion of anisotropic distribution of excited molecule axes and can be written in the form~\cite{Shternin10}:
   \begin{eqnarray}
\label{eq:D(t)}
 D^K_{\tilde q' \tilde q}(t)=\sum_n {d^{K}_{\tilde q' n}}^*\,e^{-t/\tau_r^{(n)}}\,d^{K}_{\tilde q n},
   \end{eqnarray}
   where $d^{K}_{\tilde q n}$ are are the expansion coefficients of the diffusion operator eigenfunctions over the Wigner $D$-functions and $\tau_r^{(n)}$ is a rotation diffusion time.

Equation~(\ref{eq:probe2})  was derived assuming that duration of the laser pulses is much shorter than the characteristic time of nuclear evolution and that the time of the pump pulse is $t=0$ and the time of the probe pulse is $t$.

Sum over $l$ in the first line in eq.(\ref{eq:probe2}) takes into consideration several different possible excited state lifetimes that is common for big organic and biological molecules~\cite{Lakowicz97a}. Sum over $n$ in eq.(\ref{eq:D(t)}) takes into account several possible different rotation diffusion times $\tau_{r}^{(n)}$. The number of rotation diffusion times depends on the molecular symmetry and can be as much as $(2K+1)$~\cite{Shternin10}. In case of NADH, the results of fluorescence experiments~\cite{Couprie94,Vishwasrao06,Blacker13,Vasyutinskii17} suggest that the fluorescence decay in solutions can usually be characterized satisfactory with two isotropic decay times $\tau_1$, $\tau_2$ and  a single rotation diffusion time $\tau_r$.

Transition dipole moments matrix elements in the last line in eq.(\ref{eq:probe2}) with electron and vibration wave functions shown in eq.(\ref{eq:BO approximation2}) contain integration over all electron and nuclear coordinates and over the spectral profiles of the pump and probe pulses $\Phi_{pu}(\omega)$ and $\Phi_{pr}(\omega)$. Assuming "instant" optical excitation with short laser pulses from the ground state to the first excited state PES as shown in  Fig.\ref{fig:resonant} the matrix elements of the dipole moment $\mathbf{d}^{(pu)}$ are independent of time and calculated at $t=0$. Within the model used that assumes the validity of the Born-Oppenheimer approximation the evolution time of the electron wave functions $\psi^{el}_i$ is neglected, while the evolution time of the relaxed nuclear wave functions $\chi_{1rel}$ in the matrix elements of the dipole moment $\mathbf{d}^{(pr)}$ is accounted for. Therefore we assumed that the wave functions $\chi_{1rel}$ in eq.(\ref{eq:probe2}) depend on nuclear coordinates $R$ and time $t$: $|\chi_{1rel}\rangle=|\chi_{1rel}(R,t)\rangle$ with boundary condition $|\chi_{1rel}(t=0)\rangle=|\chi_g\rangle$. The vibration wave function $\chi_{1rel}(t,R)$ can be expanded over the state 1 vibrational eigenfunctions $\chi_{v_1}(R)$ as follows:
     \begin{eqnarray}
     \label{eq:expansion}
\chi_{1rel}(R,t) = \sum_{v_1}c_{v_1}(t)\chi_{v_1}(R),
     \end{eqnarray}
where $c_{v_1}(t)$ are expansion coefficients and the indices $v_1$ label the vibrational energy states of the electronic state 1 that form the vibrational density matrix $\rho_{v_1v_1'}(t)=c_{v_1}^*c_{v'_1}$.

The diagonal elements of the density matrix $\rho_{v_1v_1'}(t)$ describe the population of the corresponding vibration energy levels during vibrational relaxation and the off-diagonal matrix elements describe quantum beats that we not observed in this experiment. Therefore, in the following we will neglect the off-diagonal density matrix elements for simplicity, however if needed they can be readily taken into account.

If the transition matrix elements in eq.(\ref{eq:probe2}) are written in the "pump" frame where the body frame axis Z is parallel to the dipole moment $\mathbf{d}^{(pu)}$, then $q_1^{\vphantom{\prime}}=q_1'=0$, $\tilde q'=0$ and the indices $q_2^{\vphantom{\prime}},q_2'$ can in general take all possible values $q_2^{\vphantom{\prime}},q_2'=0,\pm1$.

The probe transition matrix elements in eq.(\ref{eq:probe2}) can be also calculated in the "probe" frame where the body frame axis Z is parallel to the dipole moment $\mathbf{d}^{(pr)}$.  If the direction of the transition dipole moment $\mathbf{d}^{\,(pr)}$ is specified by the polar angles $\theta, \phi$ with respect to the transition dipole  moment $\mathbf{d}^{\,(pu)}$,  the $d^{\,(pr)}_{q_2}$ in eq.(\ref{eq:probe2}) is given by:
 \begin{eqnarray}
\label{eq:theta,phi}
    d^{\,(pr)}_{q_2} &=& D^{1^{\s *}}_{q_20}(\phi,\theta,0)\,d^{pr}_{0},
    \end{eqnarray}
    where $d^{pr}_{0}$ is the dipole moment component in the "probe" frame and $D^{{1}^{\s *}}_{q_20}(\phi,\theta,0)$ is a complex conjugated Wigner $D$-function.

Using eq.(\ref{eq:theta,phi}) and combining two $D$-functions and a Clebsch-Gordan coefficient  in eq.(\ref{eq:probe2}) one can readily show that:
 \begin{eqnarray}
\label{eq:product}
&&\sum_{q_2,q_2'}C^{1 q_2' }_{1 q_2 \: K \tilde q}D^{1^*}_{q_2 0}(\phi,\theta,0)D^1_{q_2'0}(\phi,\theta,0)=\nonumber \\
    &=&
C^{1 0 }_{1 0 \: K 0}
    D^K_{0 0}(\theta)=C^{1 0 }_{1 0 \: K 0}P_K(\cos\theta) .
    \end{eqnarray}
where $P_K(\cos\theta)$ is a Legendre polynomial of the order $K$=0, or 2.

The expression in eq.(\ref{eq:product}) is valid if $\tilde q =0$ and the Wigner $D$-function does not depend on the angle $\phi$ assuming that the interaction with surrounding molecules is isotropic.

After substitution of eq.(\ref{eq:expansion}) into eq.(\ref{eq:probe2}) and taken into account eq.(\ref{eq:theta,phi}) the time-dependent expression for the probe beam absorption intensity in eq.(\ref{eq:probe2}) can be presented in the form:
\begin{widetext}
\begin{eqnarray}
\label{eq:probe3}
    I_{ab}(t)&=& C_{pu}I_{pr}\sum_l\,a_l\,e^{-t/\tau_l}
   \sum_{ K} \Big(E_{K}(\textbf{e}_{pu})\cdot E_{K}(\textbf{e}_{pr})\Big)\left[C^{1 0 }_{1 0 \: K 0}\right]^2\,R_{pu}\,D^K_{0 0}(t)\,
   \langle P_2(\cos\theta(t)) \rangle,
    \end{eqnarray}
    where
    \begin{eqnarray}
\label{eq:overlap1}
\langle P_2(\cos\theta(t)) \rangle &=& \sum_{v_1,v_2}\rho_{v_1,v_1}(t)\,
|\langle\psi^{el}_2 \chi_{v_2}|d^{(pr)}_{0}| \psi^{el}_1 \chi_{v_1} \rangle|^2\,
     [P_2(\cos\theta)]_{v_2v_1}
\int\Phi_{pr}(\omega)\,\delta( \omega_{v_2}-\omega_{v_1}-\omega)\,  d\omega \,
    \end{eqnarray}
    \end{widetext}
    and
     \begin{eqnarray}
\label{eq:overlap2}
R_{pu}&=& \sum_{v_g,v_1}
|\langle\psi^{el}_1 \chi_{v_1}|d^{(pu)}_{0}| \psi^{el}_g \chi_{v_g} \rangle|^2\,N_g(v_g)\nonumber \\
&\times&
\int\Phi_{pu}(\omega)\,\delta( \omega_{v_1}-\omega_{v_g}-\omega)\,  d\omega \,
    \end{eqnarray}

The matrix elements of the dipole moment $\mathbf{d}^{pu}$ in eq.(\ref{eq:overlap2}) are written in the "pump" frame and the matrix elements of the dipole moment $\mathbf{d}^{pr}$ in eq.(\ref{eq:overlap1}) are written in the "probe" frame. The $\delta$-functions in eqs.~(\ref{eq:overlap1}) and (\ref{eq:overlap2}) indicate the validity of the energy conservation in optical transitions, they appear due to presentation of the transition dipole moment matrix elements with respect to the Fermi golden rule~\cite{Messiah63}.

The term $P_2(\cos\theta)$ in eq.(\ref{eq:overlap1}) describing the anisotropy of a two-photon process as function of the angle between the pump and probe transition dipole moments is well known and widely used for interpretation of the fluorescence~\cite{Lakowicz97a}, and pump-probe~\cite{Fleming96} experimental results. In this paper it was derived as a result of formal quantum mechanical treatment, where the angle $\theta$ depends on time $t$ as function of the nuclear configuration. The angular brackets in $\langle P_2(\cos\theta(t)) \rangle$ mean averaging over the molecular excited state vibronic distribution.

Using the particular values of the light polarization matrix $E_{KQ}(\textbf{e})$~\cite{Zare88b} in eq.(\ref{eq:probe3}) the expression for probe beam linear absorption dichroism in NADH can be presented in the usual form:
\begin{equation}
\label{eq:Delta I}
    \Delta I_{ab}(t)=I_X - I_Y = I_0 \left( a_1 e^{-t/\tau_1} + a_2 e^{-t/\tau_2} \right) r(t),
\end{equation}
where the pump light is assumed to be linearly polarized along axis X, $I_0$ is the isotropic component of the absorption intensity, $\tau_1, \tau_2$ are the lifetimes of the excited state of the molecule, and $r(t)$ is the anisotropy coefficient.

For symmetric top molecules, the coefficient  can be written in the form:
\begin{equation}
\label{eq:r}
    r(t) = \frac{2}{5} D_{00}^2(t)\langle P_2(\cos \theta(t)) \rangle.
\end{equation}

The anisotropy coefficient in eq.(\ref{eq:r}) depends on time due to contribution from two effects. The first one is rotational diffusion, which is due to dispersion of the distribution of molecular axis directions:
\begin{equation}
    D_{00}^2(t) \sim e^{-t/\tau_r}
\end{equation}
and the  second effect is due to a change in the direction of the molecular dipole moment caused by vibrational relaxation described by the time-dependence of the angle $\theta(t)$.

In general for calculating the probe beam dichroism in eq.(\ref{eq:Delta I}) having in mind eq.(\ref{eq:overlap1}) one should perform quantum chemical computations of electronic and vibrational distributions in the first three electron energy states of NADH in solution and to solve the corresponding Liouville equation for the vibration density matrix $\rho_{v_1,v_1}(t)$.

In this paper for interpretation of the experimental results we use a simple model based on the quantum relaxation theory~\cite{Blum96}.
Within this model temporal evolution of the nuclear subsystem occurs due to interaction with a thermostat containing the electron molecular subsystem and collision interactions. The model based on the Born-Oppenheimer approximation suggests that an excited electron configuration produced by absorption of a very short laser pulse "instantly" forms the molecular state 1 potential energy surface $PES_1$ and that following nuclear relaxation occurs within $PES_1$ under the action of internal interaction between molecular vibration modes and external interactions in solution.

The total Hamiltonian of the system can be written as $H= H^N + H^T + V$, where $H^N$ is the Hamiltonian of the nuclear subsystem, $H^T$ in the Hamiltonian of the thermostat, and $V$ is interaction between them. The interaction operator is: $V = V^{PES_1} + V^{int} + V^{ext}$, where $V^{PES_1}$ is the PES of the electronic state 1 calculated in a standard way within the Born-Oppenheimer approximation, $V^{int}$ is the average potential of intramolecular vibrational interactions and $V^{ext}$  is the average potential of interaction with  external molecules.

We assume that the excitation to the state 1 at $t=0$ results in vibrational wave packet described by the density matrix $\rho_{v_1v_1'}(0)$ and the following relaxation occurs as shown schematically in Fig.\ref{fig:resonant}. Within the Markovian approximation that is appropriate in our experimental conditions the relaxation of the multi-level vibration system in state 1 can be characterised by the relaxation time $\tau_{v}$ that is assumed to be much longer than fast vibrations of the molecular system. The vibrational relaxation time $\tau_v$ is determined both by the dephasing of vibrational modes and the damping due to interaction with the surrounding solvent. Within this approximation the term $\langle P_2(\cos\theta(t)) \rangle$ in eqs.(\ref{eq:overlap1}) and (\ref{eq:r}) can be presented in an approximate form:
\begin{eqnarray}
\label{eq:overlap3}
\langle P_2(\cos\theta(t)) \rangle \approx A_1+(A_0-A_1)e^{-{t/\tau_{v}}},
    \end{eqnarray}
where the time-independent terms $A_0$ and $A_1$ specify the values of the term $\langle P_2(\cos\theta(t))\rangle$ at the nuclear configurations refer to the initial molecular state $g$ at $t=0$ and state 1 (at $t\rightarrow\infty$), respectively: $A_0=\langle P_2(\cos\theta)\rangle|_{t=0}$, $A_1=\langle P_2(\cos\theta)\rangle|_{t\rightarrow\infty}$.

The developed model allows for characterisation of the dichroism signal temporal dependence with the times of rotational diffusion $\tau_r$ and vibrational relaxation $\tau_v$.

\section{Experimental section}
\subsection{Experimental approach}
\begin{figure}[h!]\center
\includegraphics[width=0.25\textwidth]{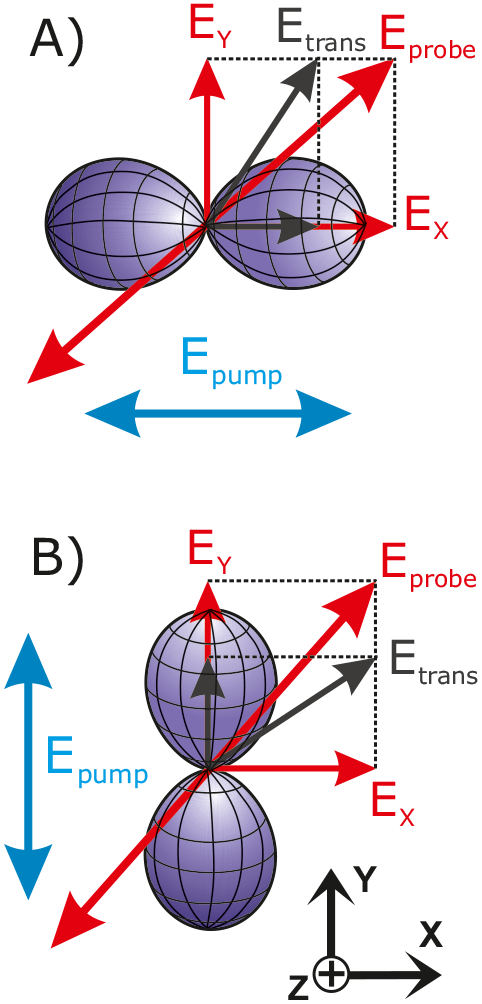}
\caption{Experimental geometry. \\
A)\,Pump beam is linearly polarized along axis X. \\
B)\,Pump beam is linearly polarized along axis Y.}
\label{fig:geometry}
\end{figure}
The pump-probe experimental geometry used in this paper  is shown schematically in Fig.\ref{fig:geometry} where the pump and probe beams
are both linearly polarized and propagate collinearly along axis Z. The duration of the pump and probe pulses was about 100~fs with a repetition rate of 80~MHz.  The pump beam at 360~nm resonantly promoted ground state NADH molecules to the first electronic excited state referred
to the nicotinamide chromophore group and produced the excited state axes/bonds alignment. The pump pulse train polarization was  modulated by a photoelastic modulator at 100 kHz  and oscillated between vertical (Y) and horizontal (X) positions shown in Figs.\ref{fig:geometry}A and \ref{fig:geometry}B.

The alignment was detected by recording of the linear dichroism $\Delta I_{ab}(t)$ of the probe beam at 720~nm that was linearly polarized on 45$^0$ to the axis Y. The probe beam transferred the molecule to the higher laying electronic excited states related to the adenine chromophore group as shown in Figs.\ref{fig:resonant} and \ref{fig:spectra}. No probe beam absorption was detected in the absence of the pump beam.

As can be seen in Fig.\ref{fig:geometry} the alignment of the excited state molecular axes produced by the pump beam led to different absorption of the $E_X$ and $E_Y$ probe beam polarization components. Therefore, the polarization of the probe beam passed through the experimental sample oscillated about its initial position at 45$^0$  with the frequency of 100 kHz with respect to the oscillation of the pump beam polarization. The difference absorption
$\Delta I_{ab}(t)$ at the frequency of 100 kHz was recorded by a sensitive balanced detection scheme.

Therefore, in the conditions of our experiment the scalar product of two light polarization matrices in eq.(\ref{eq:probe3}) oscillated as the modulation frequency $\omega=$100~kHz.  As shown in Appendix~\ref{app:B} the pump beam polarization modulated as:
 \begin{eqnarray}
\label{E:pump modul}
\mathbf{e}_{pu}(t)=\frac{1}{2}\left[\mathbf{e}_X(1+e^{i\varphi(t)})+\mathbf{e}_Y(1-e^{i\varphi(t)})\right],
\end{eqnarray}
where the phase $\varphi(t)=\pi\sin\left(\frac{\omega}{2} t\right)$ results in the following transformation of eq.(\ref{eq:Delta I}):
\begin{eqnarray}
\label{eq:Delta 2}
    \Delta I_{ab}(t)&=& I_X - I_Y  \\
    &=& I_0 \left( a_1 e^{-t/\tau_1} + a_2 e^{-t/\tau_2} \right) r(t)\cos\varphi(t).\nonumber
\end{eqnarray}

According to eq.(\ref{eq:Delta 2}) the dichroism signal was recorded in our experiments at the modulation frequency $\omega$ oscittating \emph{in quadrature} to the reference modulation signal $\sim \sin\omega t$.

Significant efforts have been undertaken to achieve the best-possible sensitivity of the method. The peculiar features of the method realization were:  high-frequency modulation of the pump pulse train \emph{polarization} at 100~kHz following  by the separation of the anisotropic contribution to the signal using a highly-sensitive balanced detection scheme, a differential integrator, and a lock-in amplifier recording the absorption change $\Delta I_{ab}(t)$ at the modulation frequency of 100~kHz in a very narrow frequency bandwidth of a few Hz. As a result, the method allowed for investigation of excited state dynamics in NADH at a less than nJ level of pump pulse energy.

\subsection{Materials}
$\alpha$-NADH (reduced nicotinamide adenine dinucleotide) disodium salt was shipped from Sigma–Aldrich. NADH was deluted in water-ethanol mixture at thirteen different volume proportions from 0 to 95\%. The ethanol purity was 96\% as determined by refractive index measurements. NADH powder was dissolved in distilled water in concentration of about 20-30 mM and then added to a previously prepared water-ethanol mixture.  NADH concentration in  solution was in all cases maintained at 0.1~mM that resulted in less than 10\% absorption of the pump light thus preventing saturation effect. Each solution were prepared fresh daily at room temperature.

\subsection{Steady-state measurements}
The absorption and fluorescence spectra of NADH (see Fig.\ref{fig:spectra}) were recorded with a spectrophotometer and spectrofluorimeter before to monitor the optical density and chemical stability of the solutions. The optical density of the solution was monitored during the entire measurement. The change in the pump beam absorption was negligible during the course of the measurements.

\subsection{Pump-probe spectroscopy measurements}
\begin{widetext}
\begin{figure}[h]\center
\includegraphics[width=0.7\textwidth]{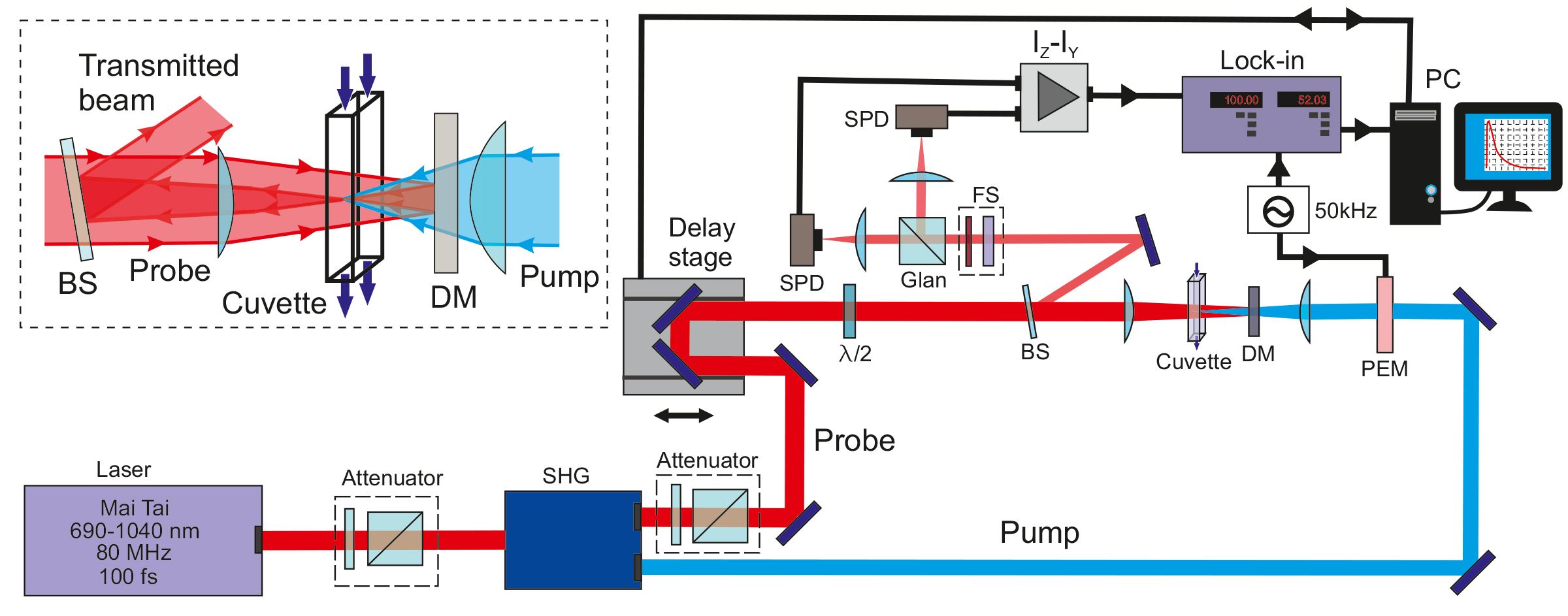}
\caption{Schematic of the experimental setup.}
\label{fig:Setup}
\end{figure}
 \end{widetext}

The schematic of the experimental setup used for our pump-probe polarization experiments that was already discussed in our recent publication~\cite{Gorbunova20} is shown in Fig.\ref{fig:Setup}.  A femtosecond Ti:Sapphire oscillator (MaiTai HP, Spectra Physics) without an optical amplifier tunable in the spectral range 720 - 800~nm was used as a light source. The laser pulse duration was about 100~fs and a 8~nm bandwidth. The pulse repetition rate was 80 MHz. The fundamental laser output with intensity of 2~W passed through an attenuator to set the pump power and then was split into pump and probe beams.

The pump beam at 360-400~nm was produced by frequency doubling by a second harmonic generator (SHG) (Inspire Blue, Spectra Physics). After SHG the pump beam travelled through a photoelastic modulator (PEM) (PEM-100, Hinds Instruments) operating at the frequency of 50~kHz, that periodically switched the direction of the pump pulse train polarization plane from vertical (Y) to horizontal (X) positions at the frequency of 100~kHz. After PEM the pump beam was focused by a lens through a dichroic mirror (DM) onto the center of a 1~mm passlength quartz cuvette to a 10~$\mu$m spot size. NADH solution circulated through the cuvette by a peristaltic pump at a flow speed of 0.2~m/s to prevent NADH photobleaching. The energy of the pump pulses entering the cuvette was about 2~nJ.
The fundamental output of the laser at 720-800~nm was used as a probe beam. The probe beam passed successively through an attenuator and a motorized delay stage producing the probe pulse delay of the time $t$ with respect to the pump pulse. The probe light polarization plane was fixed at 45$^\circ$ to Y axis by a zero-order $\lambda/2$ phase plate and then passed through a beamsplitter (BS) and was focused by a lens. As shown in inset in Fig.\ref{fig:Setup} the probe beam passed through the absorption cuvette and then was reflected back by DM and focused onto the cuvette center. The probe beam spot size on the cuvette was about 20~$\mu$m. The focal regions of the beams overlapped considerably only when the probe beam was reflected back from the DM, therefore the beams interacted with NADH molecules mainly when they propagated inside the absorption cuvette in parallel to each other. The probe pulses energy on the cuvette was about 0.5~nJ.

The transient probe beam exited the cuvette was reflected by a BS and then directed to a Glan prism (Glan) that separated two orthogonally polarized polarization components $I_X$ and $I_Y$. To avoid the pump light scattering into the detection channel, a filtration system (FS) consisted of a dichroic
mirror and an absorption filter was placed in front of Glan.

The probe beam polarization components $I_X$ and $I_Y$ separated by Glan were recorded by a balanced detection scheme consisted of two identical silicon photodiodes (SPD) (DET10A/M, ThorLabs) and a differential integrator (DI) with a passband of 0-–4~MHz. The balanced scheme was adjusted to have zero output electric signal when the pump beam was tuned off and therefore effectively cancelled fluctuations in the probe beam amplitude and in the solution density.
The output differential signals of DI amplitude-modulated at 100~kHz were fed to a lock-in amplifier (LOCK-IN) (SR844 RF, Stanford Instruments) with a narrow frequency bandwidth of 3 Hz and then collected and analysed by a computer.

\subsection{Polarised fluorescence spectroscopy measurements}
The method of polarized fluorescence spectroscopy was also used in this paper for complementary determination of the isotropic delay times $\tau_1$ and $\tau_2$, corresponding weighting coefficients $a_1$ and $a_2$, and the rotation diffusion time $\tau_r$ at large time scales. These experiments were carried out with the experimental setup discussed in detail in our recent publications~\cite{Herbrich15,Sasin18}.

\section{Experimental results}
\subsection{Experimental signals}
\begin{figure}[h!]\center
\includegraphics[width=0.45\textwidth]{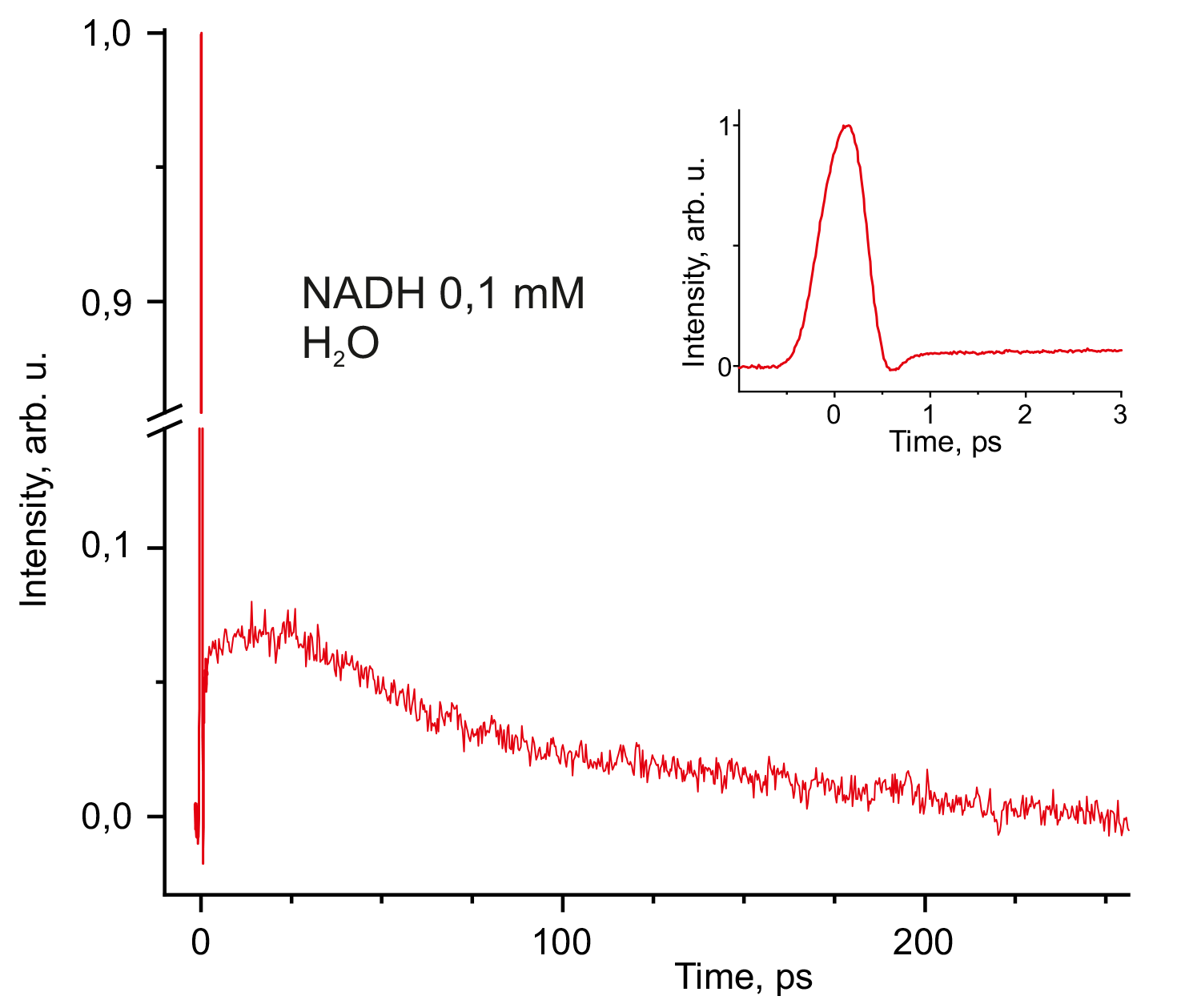}
\caption{Typical experimental signal in NADH in aqueous solution}
\label{fig:NADH}
\end{figure}

A typical experimental signal observed in  NADH  in water solution at 0.1~mM concentration is shown in Fig.~\ref{fig:NADH}. As can be seen the signal contains a high narrow peak at relatively small delay times less than $t \simeq 1~ps $ with a wing having a much wider and smaller maximum at lager delay times. The narrow peak is shown in the inset in Fig.~\ref{fig:NADH} in an extended scale.

\begin{figure}[h!]\center
\includegraphics[width=0.45\textwidth]{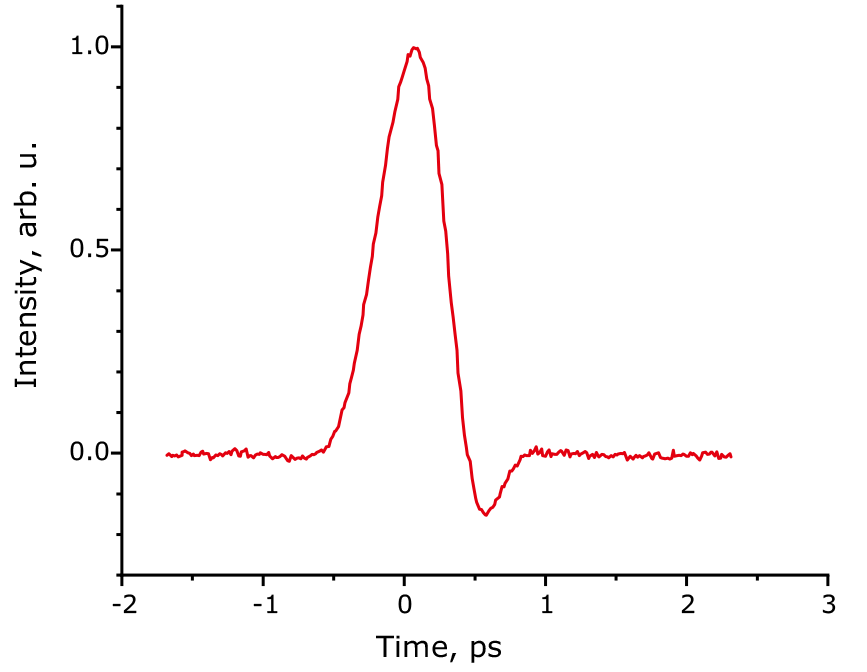}
\caption{Experimental signal in pure water}
\label{fig:Water}
\end{figure}

Similar peaks were also observed in pure water and all water-ethanol solutions did not containing NADH. The corresponding peak recorded in water is shown in Fig.\ref{fig:Water}. As can be seen in Fig.\ref{fig:Water} and was proved in our experiments in ethanol-containing solutions, these peaks contain no longer-time wing, the wing was observed in NADH-containing solutions only.

We suggest that the peak reflects a complicated cross-phase modulation dynamics occurring in water and other liquid phases during the
time of overlap between the pump and probe pulses similar to that reported elsewhere~\cite{Miller94,Wiersma01}. The study of these effects is out of the scope of this paper.

At larger delay times (approximately $t \geq 1$ ps) the signal in Fig.\ref{fig:NADH} reflects the dynamics of anisotropic vibrational and rotational
relaxation in NADH.  The interpretation of the shape of this part of the signal in terms of linear dichroism of the probe beam is presented in Sec.\ref{sec:theory}. With respect to the model developed in Sec.\ref{sec:theory} the growing part of the anisotropy signal in Fig.\ref{fig:NADH} at $t \leq 25$~ps is mostly reflects vibration relaxation in the first electronic excited state in NADH resulting in rotation of the molecular transition dipole moment, and the decaying part of the anisotropy signal at $t \geq 25$ ps can be explained by  rotation diffusion of excited NADH molecules in solution.

\begin{figure}[h!]\center
\includegraphics[width=0.4\textwidth]{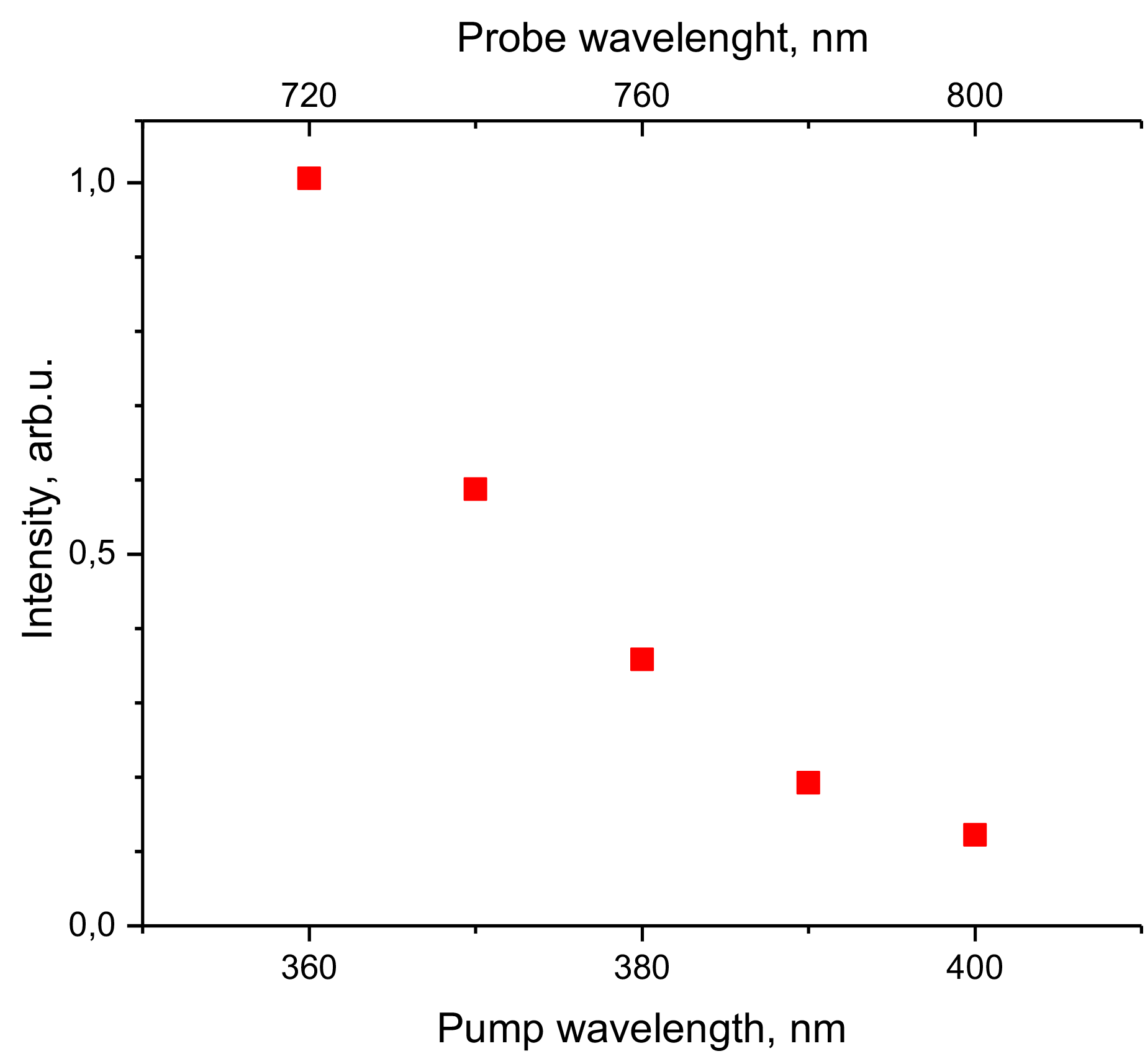}
\caption{Signal in NADH as function of the pump beam wavelength}
\label{fig:lambda}
\end{figure}

For proving the resonance behavior of the anisotropic signal in NADH we investigated the maximum signal amplitude at $25$~ps as function of the pump beam wavelength. The result obtained is shown in Fig.\ref{fig:lambda}. As can be seen in Fig.\ref{fig:lambda} the signal amplitude decreases dramatically when the pump beam wavelength increases from $\lambda$=360~nm to $\lambda$=400~nm.  This observation is in a good agreement with the resonance excitation scheme in Fig.\ref{fig:resonant} and the NADH absorption band shown in Fig.\ref{fig:spectra}.

\begin{figure}[h!]\center
\includegraphics[width=0.4\textwidth]{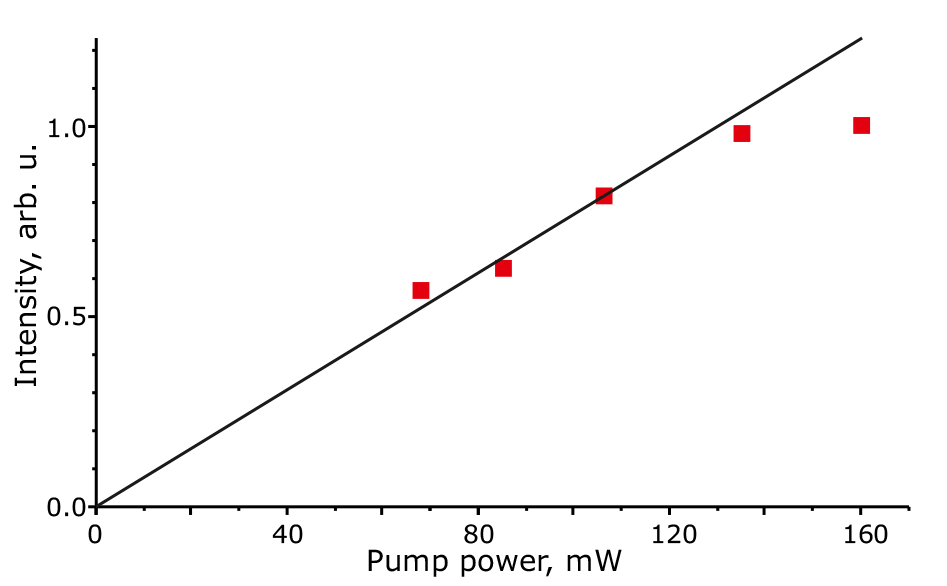}
\caption{Signal in NADH as function of the pump beam intensity}
\label{fig:Intensity}
\end{figure}

The dependence of the signal maximum amplitude on the pump beam intensity is presented in Fig.\ref{fig:Intensity}.  As can be seen, at relatively small pump beam intensities the signal amplitude increases linearly with pump intensity as expected for a one-photon absorption and only at large intensities the signal deviates from the linear behaviour manifesting saturation.

\subsection{Experimental data processing}
According to eqs.(\ref{eq:Delta I}), (\ref{eq:r}), and (\ref{eq:overlap3}) the experimental pump-probe signal $\Delta I_{ab}(t)$ can be described by the expression:
\begin{equation}
    \Delta I_{ab}(t) = (A_1 + (A_0 - A_1)e^{-t/\tau_v})e^{-t/\tau_{rot}}w(t), \label{eq:fitpp}
\end{equation}
where the lifetime decay $w(t)$ has a form
\begin{equation}
    w(t) = a_1 e^{-t/\tau_1} + a_2 e^{-t/\tau_2}.
\end{equation}

Using eq.(\ref{eq:fitpp}) we determined  the best estimation for vibrational relaxation time $\tau_v$, rotational diffusion time $\tau_{r}$, and coefficients $A_0, A_1$ by fit as function of the ethanol concentration in the sample, while the lifetimes $\tau_1, \tau_2$ and the weighting coefficients $a_1, a_2$ were determined independently in polarized fluorescence spectroscopy experiments.  The experimental values of the lifetimes $\tau_1, \tau_2$  and the ratio of the coefficient $a_1, a_2$  are collected in Table~\ref{tab:fluo}.
\begin{table}[h]
\caption{\label{tab:fluo} Lifetimes $\tau_1$, $\tau_2$, and the ration of the weighting coefficients $a_1/a_1$ in NADH at different ethanol concentrations determined from polarization fluorescence experiments.}
\begin{center}
\begin{tabular}{ | c | c | c | c |}
\hline
EtOH, \% vol & $\tau_1$, ps & $\tau_2$, ps & $a2/a1$ \\ \hline
0 \% & 230 $\pm$ 40 & 590 $\pm$ 50 & 0.34 $\pm$ 0.02\\ \hline
16 \% & 245 $\pm$ 35 & 600 $\pm$ 40 & 0.38 $\pm$ 0.02 \\ \hline
27\% & 280 $\pm$  30 &  680 $\pm$ 40 & 0.31 $\pm$ 0.01\\ \hline
45\% & 280 $\pm$ 20& 670 $\pm$ 10 &  0.43 $\pm$ 0.03\\ \hline
64\% & 300 $\pm$ 30 & 720 $\pm$ 30 & 0.48 $\pm$ 0.07 \\ \hline
92\% & 300 $\pm$ 30 & 830 $\pm$ 20 & 0.71 $\pm$ 0.14 \\
\hline
\end{tabular}
\end{center}
\end{table}

\begin{figure}[h!]
    \includegraphics[width=0.4\textwidth]{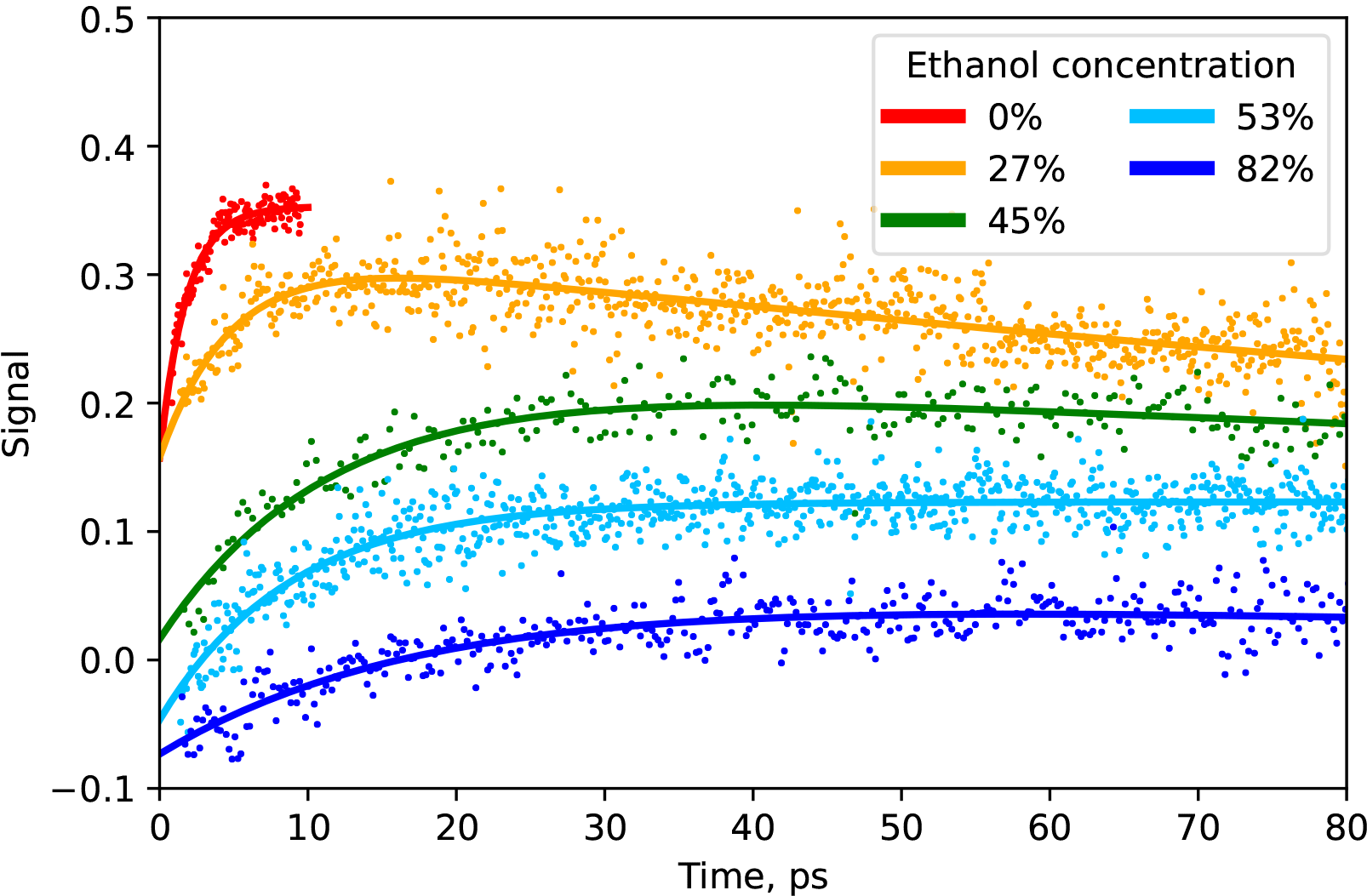}
    \caption{Pump-probe signals at various ethanol concentrations at small timescale. \\
    Dots are experimental data and solid lines are fit.}
    \label{fig:short_fits}
\end{figure}

\begin{figure}[h!]
    \includegraphics[width=0.4\textwidth]{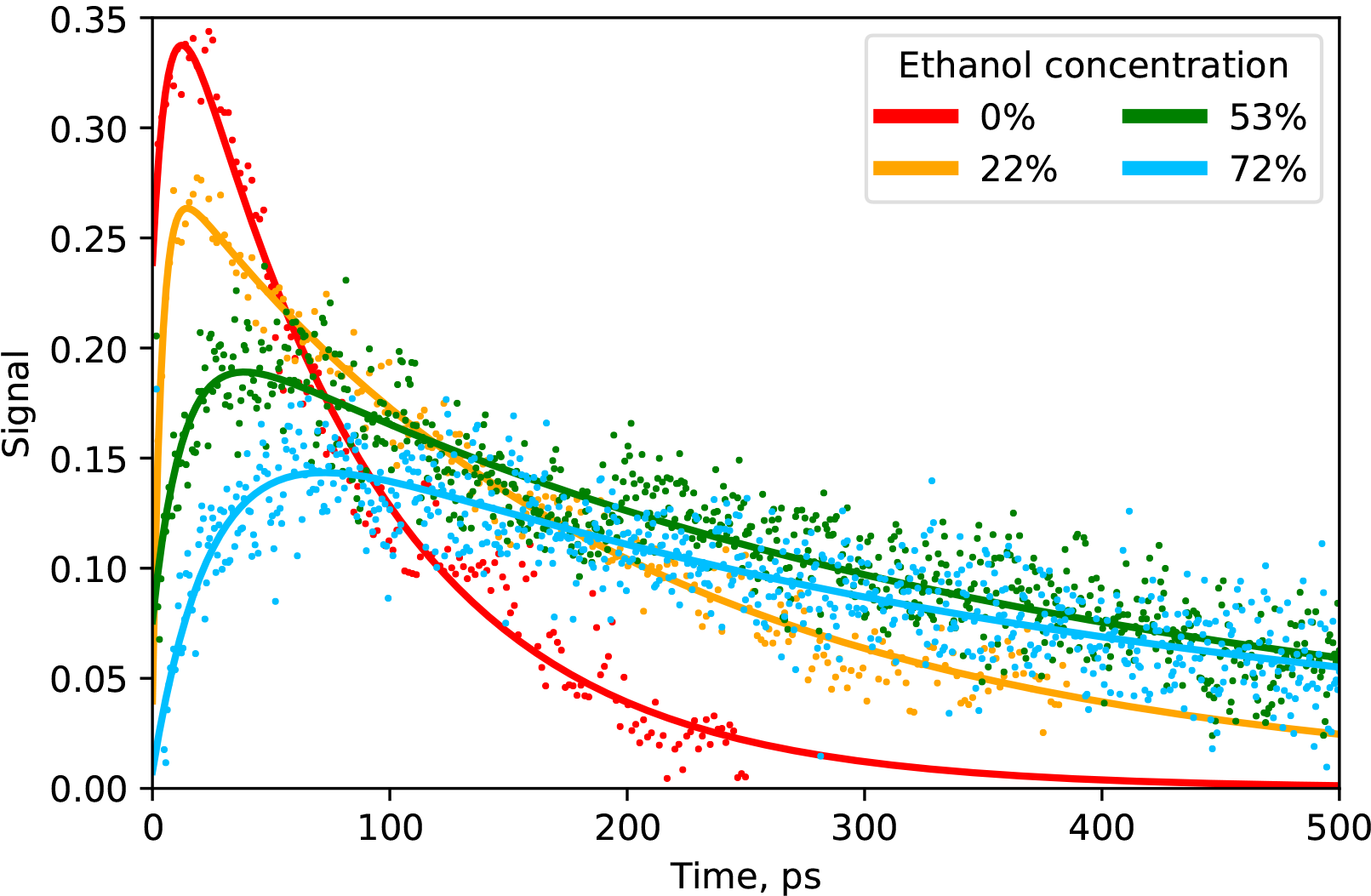}
    \caption{Pump-probe signals at various ethanol concentrations at large timescale. \\
    Dots are experimental data and solid lines are fit.}
    \label{fig:long_fits}
\end{figure}

The experimental signals $\Delta I_{ab}$ recorded at different ethanol concentrations are presented in Figs.~\ref{fig:long_fits} and \ref{fig:short_fits} at two different timescales, where the dots are our experimental data and solid lines are our fit.

The noise in the measured signal can be assumed as an additive Gaussian noise with equal variance. Therefore, we use a simple objective function $\Phi = \sum_i(\Delta I_{ab}(t_i) - (\Delta I_{ab})_i)^2$ to find fitting parameters $A_0, A_1,\tau_v$, and $\tau_{r}$. For solving the nonlinear optimization problem we used differential evolution to find the global minimum of the objective function $\Phi$ \cite{Storn1997}. Unlike local optimization schemes, this procedure is robust to an initial guess of the fitting parameters.

Experimental data in Figs.~\ref{fig:short_fits} and  \ref{fig:long_fits} manifest rapid increase of the signal due to vibrational relaxation and slow decay due to rotational diffusion. Since the characteristic times $\tau_{r}$  were found to be much longer than $\tau_v$, for accurate determination of $\tau_v$ we used shorter delays $t$ in Fig.~\ref{fig:short_fits} and for determination of $\tau_{r}$ we used longer delays in Fig.~\ref{fig:long_fits}.


\section{Discussion}
\begin{figure}[h!]
    \includegraphics[width=0.4\textwidth]{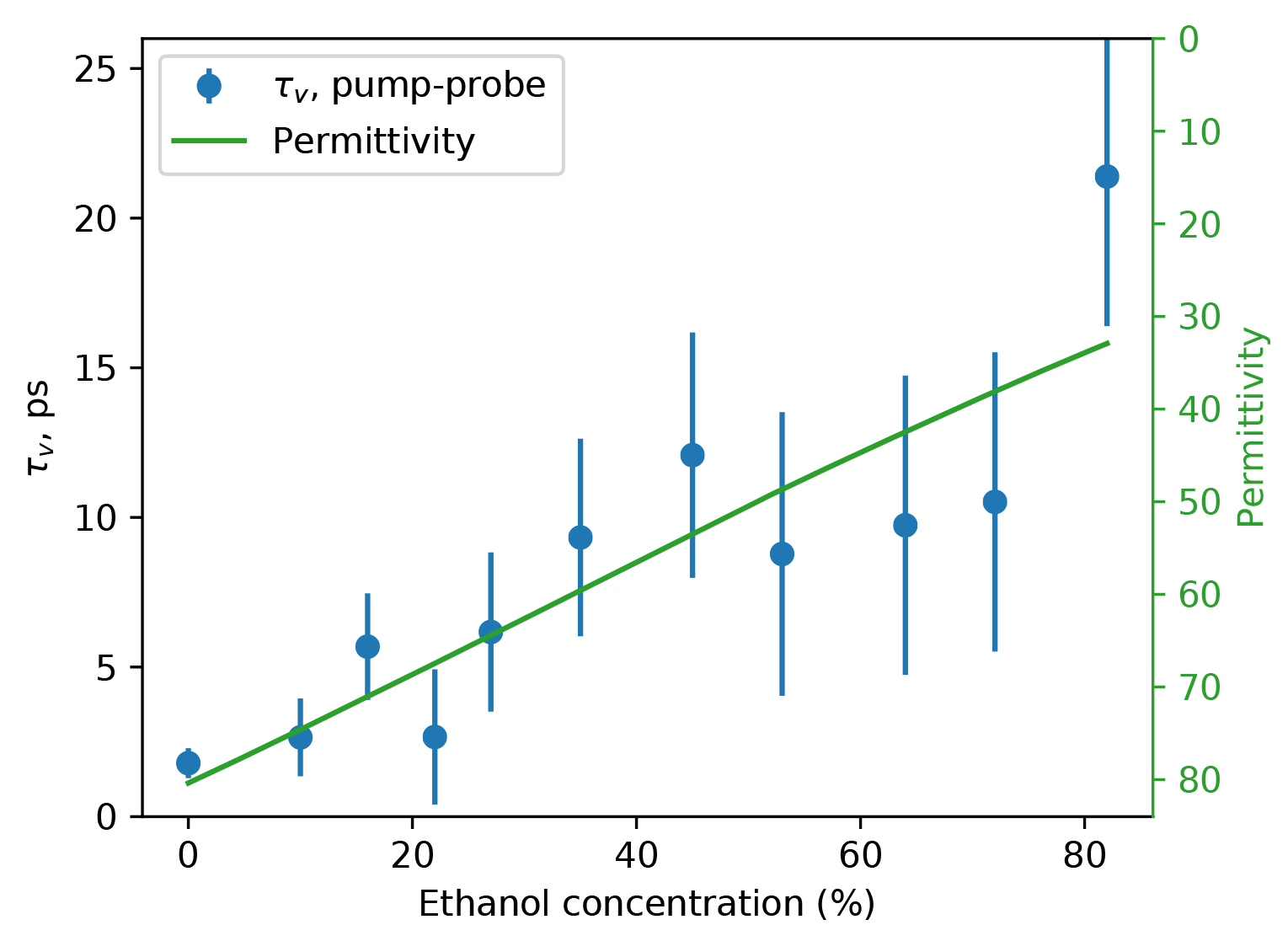}
    \caption{Vibrational relaxation time $\tau_v$ in the first excited state of NADH determined from experiment as function of ethanol concentration in water-ethanol solution.\\ The solid line is solution permittivity. }
    \label{fig:EtOH_tau_vib2}
\end{figure}

The relaxation time  $\tau_v$ in eqs.~(\ref{eq:Delta I})-(\ref{eq:overlap3}) and (\ref{eq:fitpp}) characterizes the velocity of vibrational relaxation in NADH first excited state schematically shown in Fig.~\ref{fig:resonant}. The $\tau_v$ value determined from experiment as function of ethanol concentration in water-ethanol solution is given in Fig.~\ref{fig:EtOH_tau_vib2}. As can be seen in Fig.~\ref{fig:EtOH_tau_vib2}, $\tau_v$  increases linearly with ethanol concentration within the experimental error bars. This relationship can be attributed to interaction of excited state NADH with surrounding solution molecules. Water molecules are more polar than ethanol and therefore more strongly perturb NADH wave functions. Therefore NADH-water interactions leads to faster vibrational relaxation and to stronger influence on NADH nuclear configuration~\cite{Hull01, Smith00}.

The solvent static permittivity of water-ethanol solution~\cite{Akerlof32} as a relative measure of chemical polarity is plotted in Fig.~\ref{fig:EtOH_tau_vib2} with a solid straight green line in an inverse scale given in the right. As can be seen in Fig.~\ref{fig:EtOH_tau_vib2} the vibration relaxation time $\tau_v$ is inverse proportional to the solution permittivity within the experimental error bars.

According to the results of the theory given in Sec.~\ref{sec:theory} the coefficients $A_0$ and $A_1$ in eqs.(\ref{eq:Delta I})-(\ref{eq:overlap3}) have important physical meaning being the limiting values of the term  $\langle P_2(\cos\theta(t))\rangle$, where $\theta$ is the angle between the directions of dipole moments $\mathbf{d}^{(pu)}$ and $\mathbf{d}^{(pr)}$ on the optical transitions $g \rightarrow 1$ and $1 \rightarrow 2$ in Fig.~\ref{fig:resonant}, respectively. Therefore the coefficients $A_0$ and $A_1$ characterise the rotation of the molecular transition dipole moment during excited state vibrational relaxation. The coefficient $A_0$ is equal to $\langle P_2(\cos(\theta))\rangle|_{t=0}$ at the time of the pump pulse excitation where the angle $\theta$ refers to the ground state equilibrium nuclear configuration for both $g \rightarrow 1$ and $1 \rightarrow 2$ optical transitions. In the course of vibrational relaxation the excited state nuclear configuration of NADH transforms to a new equilibrium position schematically indicated as the minimum on PES 1 in Fig.~\ref{fig:resonant}. The coefficient $A_1$ is equal to $\langle P_2(\cos(\theta))\rangle|_{t\rightarrow\infty}$  where the angle $\theta$ refers to the ground state equilibrium nuclear configuration for $g \rightarrow 1$ optical transition and to the excited state 1 nuclear configuration for the $1 \rightarrow 2$ optical transition.

\begin{figure}[h!]
    \includegraphics[width=0.4\textwidth]{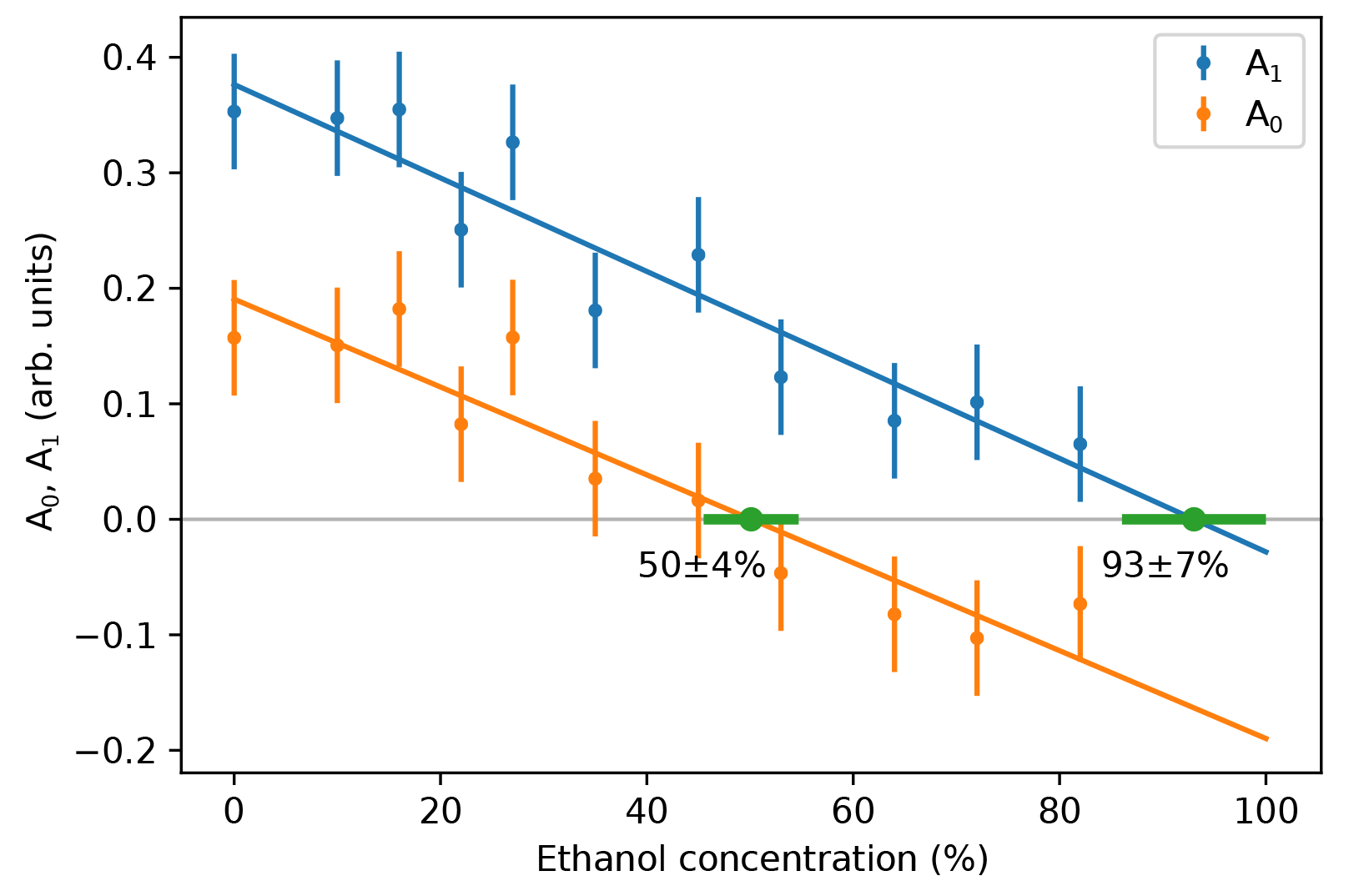}
    \caption{Coefficients $A_0$ and $A_1$ as function of ethanol concentrations in water-ethanol solution.\\
    Solid lines are linear fit.}
    \label{fig:pp_coeff}
\end{figure}

The coefficients $A_0$ and $A_1$ as function of ethanol concentrations are plotted in Fig.~\ref{fig:pp_coeff}.

According to the definitions below eq.(\ref{eq:overlap3}) the coefficients may in general have either positive, or negative values in the range $1 \geq A_{0,1} \geq -1/2$ where the maximum value $A_{0,1}=1$ refers to the case when the transition moments $\mathbf{d}^{(pu)}$ and $\mathbf{d}^{(pr)}$  on the transitions  $g \rightarrow 1$ and $1 \rightarrow 2$ are parallel to each other and the minimum value $A_{0,1}=-1/2$ refers to the case when these transitions are perpendicular to each other.

One can conclude from Fig.~\ref{fig:pp_coeff} that the coefficient $A_1$  is always larger than the coefficient $A_0$. This result can be interpreted within two scenarios that may be complementary to each other: (i) the angle $\theta$ between the transition dipole moments $\mathbf{d}^{(pu)}$ and $\mathbf{d}^{(pr)}$ is larger at time $t=0$ than at time $t\rightarrow\infty$ and (ii) the overlap vibrational integrals and/or modulus of the dipole moments $\mathbf{d}^{(pr)}$ and $\mathbf{d}^{(pu)}$ in eqs.(\ref{eq:overlap1}) and (\ref{eq:overlap2}) are greater for the probe optical transition than for the pump one. The mechanisms (i) and (ii) cannot be isolated with the methods used in this paper, the isolation needs the assistance of intensive quantum chemical computations.

As can be seen in Fig.~\ref{fig:pp_coeff} both coefficients decrease linearly with ethanol concentration within the experimental error bars and at certain concentrations change their sign.  This behavior of the coefficients $A_0$ and $A_1$ suggests that the angle $\theta$ increases with ethanol concentration due to change of equilibrium nuclear configurations as function of the solution polarity. The coefficients $A_0$ and $A_1$ change their sign at the magic angle $\theta_M=54.7^\circ$. According to Fig.~\ref{fig:pp_coeff}  this point is achieved by the coefficient $A_0$ at about  50$\pm$4\% ethanol concentration and by the coefficient $A_1$ at 93$\pm$7\% ethanol concentration.

\begin{figure}[h!]
    \includegraphics[width=0.4\textwidth]{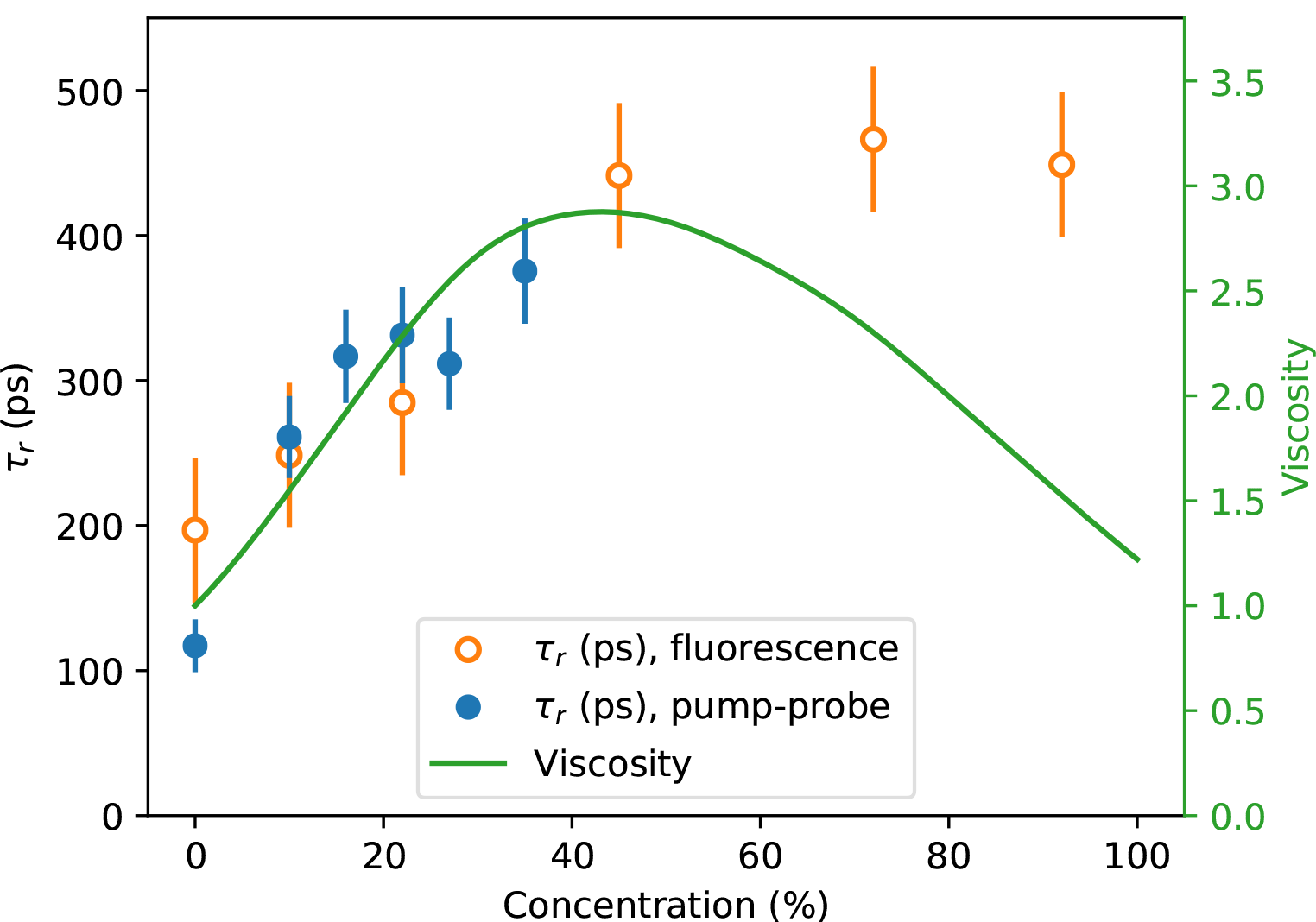}
    \caption{Rotational diffusion time $\tau_r$ as a function of ethanol concentrations in water-ethanol solution. \\
    Symbols are experimental data and the solid curve is solution viscosity. }
        \label{fig:EtOH_tau_rot}
\end{figure}

The rotation diffusion time  $\tau_r$ in eqs.~(\ref{eq:Delta I})-(\ref{eq:overlap3}) and (\ref{eq:fitpp}) characterizes the blurring of the distribution of excited NADH axes due to interaction with surrounding solvent molecules. The rotation diffusion time determined by the polarization-modulation transient and by the fluorescence polarization method as a function of ethanol concentration in water-ethanol solution is given in Fig.~\ref{fig:EtOH_tau_rot} with symbols.

In contrast to $\tau_v$, the dependence of $\tau_{r}$ on ethanol concentration has a non-linear behavior. As reported by Millar et al~\cite{ZEWAIL79} and by Potma et al~\cite{Wiersma01} in certain liquids the rotation diffusion time is directly proportional to the solution viscosity.

Our results show that in water-ethanol solution it is not exactly the case. For comparison, the water-ethanol solution viscosity~\cite{Jouyban12} is plotted in Fig.~\ref{fig:EtOH_tau_rot} with a solid red curve with a maximum. As can be seen in Fig.~\ref{fig:EtOH_tau_rot} the rotation diffusion time $\tau_{r}$ is proportional to the solution viscosity at the methanol concentrations less than 40\% only, while at higher concentrations deviation  from the direct proportionality is observed. This behavior can be attributed to the change of the NADH nuclear configuration and redistribution of different NADH conformers population at high ethanol concentrations.

%\clearpage

\section{Conclusions}
The paper presents the results of the study of fast anisotropic relaxation and rotational diffusion in the first electronic excited state of biological coenzyme NADH in water-ethanol solutions at various ethanol relative concentrations. Novel polarization-modulation transient method has been used for detailed study of anisotropic vibrational relaxation. The method combines high-frequency modulation of the pump pulse train {polarization} at 100~kHz following  by the separation of the anisotropic contribution to the signal using a highly-sensitive balanced detection scheme, a differential integrator, and a lock-in amplifier recording the linear dichroism of the probe beam at the modulation frequency in a narrow frequency bandwidth of a few Hz. As a result, the method allows for investigation of sub-picosecond excited state dynamics at a less than nJ level of pump pulse energy. Interpretation of the experimental results was done by a model expressions based on the full quantum mechanical treatment using the state multipole representation of the molecular density matrix. It was shown that the dynamics of anisotropic relaxation in NADH under excitation with 100~fs pump laser pulses can be characterised by a single vibration relaxation time $\tau_v$ laying in the range 2--15~ps and a single rotation diffusion time $\tau_r$ laying in the range 100--450~ps. The vibration relaxation time  $\tau_v$ was found to be proportional to the solvent polarity in the entire range of ethanol concentrations. The rotation diffusion time $\tau_r$ was found to be proportional on the solvent viscosity at the ethanol concentrations less then 40\%, however at higher concentration the deviation from the proportionality was observed. The rotation of the transition dipole moment in the course of vibrational relaxation has been analyzed by exploring the experimental parameter $\langle P_2(\cos\theta(t))\rangle$.

\section{Conflict of interests}
There are no conflicts to declare.

\acknowledgments
The study was supported by the Russian Foundation for Basic Researches under the grant No 18-03-00038. YMB, AAS, and OSV are grateful to BASIS Foundation for financial support under the grant No 19-1-1-13-1. The authors are grateful to the Ioffe Institute for providing the equipment used in the experiments.

\begin{appendix}
\section{Derivation of eq.(\ref{eq:probe2}).}
\label{app:A}
In the first order of the perturbation theory the probe beam absorption intensity $I_{ab}$ in the optical transition $1 \rightarrow 2$ can be described by the expression~\cite{Blum96}:
\begin{eqnarray}
\label{eq:probe}
    I_{ab}(t)&=& I_{pr}\sum_{J_2,M_2}\sum_{J_1M_1,J_1'M_1'}\Phi_{pr}\,
    \langle J_{2}M_{2}|\,V_{pr} |\,J_1 M_1\rangle^*\nonumber \\
    &\times&
    \langle J_{2}M_{2}\,|\,V_{pr} |\,J_1'M_1'\rangle \,\rho_{\sss J_1' M_1',J_1 M_1}(t),
\end{eqnarray}
where $I_{pr}$ is the incident probe beam intensity, $ \rho_{\sss J_1' M_1',J_1 M_1}(t)$ is the molecular first excited state density matrix, $\Phi_{pr}$ is a spectral profile of the probe pulse, and the terms in angular brackets are transition dipole moment matrix elements.

We assume that the density matrix $  \rho_{\sss J_1' M_1',J_1 M_1}(t)$ in eq.(\ref{eq:probe}) depends on time $t$ due to energy transfer and relaxation processes in the excited state. In general, the density matrix $\rho_{\sss J M' J M}$  in eq.~(\ref{eq:probe}) depends also on other quantum numbers which are dropped for brevity.

If the duration of the excitation laser pulse at $t=0$ is much shorter than the characteristic time of excited state vibrational relaxation the expression for the first excited state density matrix can be written in the form:
\begin{eqnarray}
\label{eq:pump}
    \rho_{\sss J_1' M_1',J_1 M_1}(0)&=& C_{pu}\sum_{J_g,M_g}\Phi_{pu}\,
    \langle J_{1}M_{1}|\,V_{pu} |\,J_g M_g\rangle^*\nonumber \\
    &\times&
    \langle J'_{1}M'_{1}\,|\,V_{pu} |\,J_g M_g\rangle \,N(J_g,k_g),
\end{eqnarray}
where $\Phi_{pu}$ is a spectral profile of the pump pulse and $N(J_g,k_g)$ is the population of the distribution of the ground state energy levels that is assumed to be isotropic.

We formulate eqs.~(\ref{eq:probe}) and (\ref{eq:pump}) in terms of the covariant state multipole components of the density matrix:
\begin{eqnarray}
\label{eq:statemult2}
\rho_{\sss KQ}&=& \sum_{M,M'} (-1)^{J-M'} \,C^{K Q }_{J M \: J -M'}\,
\,
\rho_{\sss J M' J M},
\end{eqnarray}

The inverse transformation is given by:
\begin{eqnarray}
\label{eq:statemult3} \rho_{\sss J M' J M}&=&\sum_{\sss KQ} (-1)^{J-M'} \,C^{K Q }_{J M \: J -M'}\,
\rho_{\sss KQ},
\end{eqnarray}
where $K$ and $Q$ are the rank ($K=0...2J$) and its component on the laboratory frame axis Z ($Q=-J...J$), respectively~\cite{Zare88b}.

Combining eqs.~(\ref{eq:probe})-(\ref{eq:statemult3}) and proceeding the transformations of the angular momentum algebra similar to that used in refs.~\cite{Denicke10,Shternin10} one comes to eq.(\ref{eq:probe2}).

\section{Derivation of eq.(\ref{eq:Delta 2}).}
\label{app:B}
Assuming that the pump beam polarization is modulated by an photo-acoustic modulator according to eq.(\ref{E:pump modul}), where $\varphi(t)=\pi\sin\left(\frac{\omega}{2} t\right)$ we calculate the spherical pump light components:
\begin{eqnarray}
\label{E:pump 1}
\mathbf{e}_{1}(t)=-\frac{1}{2\sqrt2}[\mathbf{e}_{X}(1+e^{i\varphi(t)})+
i\mathbf{e}_{Y}(1-e^{i\varphi(t)})], \\
\mathbf{e}_{-1}(t)=\frac{1}{2\sqrt2}[\mathbf{e}_{X}(1+e^{i\varphi(t)})-
i\mathbf{e}_{Y}(1-e^{i\varphi(t)})]
\end{eqnarray}
and then consider the matrix elements of the pump light polarization matrix $E_{KQ}$ with respect to eq.(\ref{E:Edef}).
 \begin{eqnarray}
\label{eq:E00}
E_{00}&=&-\frac{1}{\sqrt3},  \\
\label{eq:E20}
E_{20}&=&-\frac{1}{\sqrt6}, \\
\label{eq:E22}
E_{2\pm2}&=&\frac{1}{2}\cos\varphi(t).
\end{eqnarray}
\\

As can be seen in eds.(\ref{eq:E00})-(\ref{eq:E22}) only the pump light matrix elements $E_{22}$ and $E_{2-2}$ are modulated in time and can contribute to the experimental signal.

Substituting eds.(\ref{eq:E00})-(\ref{eq:E22}) and the probe light polarization matrix elements related to $X$ and $Y$ polarization  components~\cite{Zare88b} to the scalar product in eq.(\ref{eq:probe3}) the linear dichroism signal cane be presented in the form:
\begin{eqnarray}
\label{eq:fluorescence fin}
I_{X}(t)-I_{Y}(t)
&\sim&
 \Big[1 + \left(\frac{1}{2}+\frac{3}{2}\cos[\varphi(t)]\right) \,r(t)\Big] \nonumber \\
&-&
 \Big[1 + \left(\frac{1}{2}-\frac{3}{2}\cos[\varphi(t)]\right) \,r(t)\Big]\nonumber \\
&=&
3\cos[\varphi(t)] \,r(t),
\end{eqnarray}
where the anisotropy $r(t)$ is given in eq.(\ref{eq:r}).

Expanding the function $\cos[\varphi(t)]$ into the Fourier series one can readily prove that it oscillates in quadrature to the doubled frequency modulation signal $\sin[\omega t]$.

 \end{appendix}

%\bigskip
%\noindent Add citations manually or use BibTeX. See \cite{Zhang:14,OSA,FORSTER2007,testthesis,manga_rao_single_2007}.

%Bibliography
\bibliographystyle{unsrt}
\bibliography{pumpprobe}

%\begin{thebibliography}{100}

%\bibitem{ref1} Reference 1.

%\bibitem{ref2} Reference 2.

%\end{thebibliography}

\end{document}